\documentclass{mem}
\usepackage{natbib}\usepackage{txfonts}\usepackage{balance}
\usepackage{graphicx}
\usepackage[a4paper]{hyperref}
\idline{75}{282}
\begin{document}
\def\teff{$T\rm_{eff }$}
\def\kms{$\mathrm {km s}^{-1}$}

\title{
Jet-BLR connection in the radio galaxy 3C\,390.3}

   \subtitle{}

\author{
T.\ G. \,Arshakian\inst{1},\, J. \, Le\'on-Tavares\inst{1,2},\, A.\ P.
\,Lobanov\inst{1},\, V.\ H. \,Chavushyan\inst{2}, L.
\,Popovic\inst{3}, A.\ I. \,Shapovalova\inst{4}, A.
\,Burenkov\inst{4} and J.\ A. \,Zensus\inst{1}
          }

  \offprints{T.G. Arshakian}

\institute{ MPI f\"ur Radioastronomie, Auf dem H\"ugel 69, 53121
Bonn, Germany \email{tarshakian@mpifr-bonn.mpg.de} \and Instituto
Nacional de Astrof\'{\i}sica \'Optica y Electr\'onica, Apartado
Postal 51 y 216, 72000 Puebla, Pue, M\'exico \and Astronomical
Observatory, Volgina 7, 11160 Belgrade, Serbia \and Special
Astrophysical Observatory of the Russian AS, Nizhnij Arkhyz,
Karachaevo-Cherkesia 369167, Russia
 }

\authorrunning{Arshakian et al.}

\titlerunning{Jet-BLR connection}

\abstract{ Variations of the optical continuum emission in the radio
galaxy 3C\,390.3 are compared to the properties of radio emission
from the compact, sub-parsec-scale jet in this object. We showed
that \emph{very long-term variations} of optical continuum emission
($\ga 10$ years) is correlated with the radio emission from the base
of the jet located above the disk, while the optical \emph{long-term
variations} (1-2 years) follows the radio flares from the stationary
component in the jet with time delay of about 1 year. This
stationary feature is most likely to be a standing shock formed in
the continuous relativistic flow seen at a distance of $\sim 0.4$
parsecs from the base of the jet. To account for the correlations
observed we propose a model of the nuclear region of 3C\,390.3 in
which the beamed continuum emission from the jet and counterjet
ionizes material in a subrelativistic outflow surrounding the jet.
This results in the formation of two conical regions with
double-peaked broad emission lines (in addition to the conventional
broad line region around the central nucleus) at a distance
$\approx$ 0.6 parsecs from the central engine.
  }
\maketitle{}

\section{Introduction}
Active galactic nuclei (AGN) are believed to be powered by accretion
of a disk material into the central nucleus (or black hole;
hereafter, BH). The accretion energy is transformed to the radiation
of \emph{thermal} variable continuum emission generated in the disk
and to the kinetic energy of the bipolar outflows of plasma material
ejected in the directions normal to plane of the disk. In radio-loud
AGN, the outflows (jets) are collimated and highly relativistic, and
\emph{synchrotron} continuum emission from jets may dominate at all
energies \citep{arshakian_ulrich97,arshakian_worrall05}. The continuum variable emission
in these AGN may be related to both the jet and accretion flows and,
hence, to be a mixture of thermal and non-thermal emission. It is
believed that continuum radiation ionizes the clouds in the
broad-line regions (BLR), therefore the localization of the source
of variable continuum emission is instrumental for understanding the
structure of the central engine and its operation. The jets
originate very near to the BH and may trace the environment of an
AGN from subpc- to kpc-scales in directions of advancing.

To localize the source of optical continuum emission in the jet we
search for a positive correlation between the variability of the
optical continuum flux and radio flux density of the parsec-scale
jet for the radio-loud galaxy 3C\,390.3 ($z=0.0561$). We combined
archived monitoring data of $\ga100$ optical data points with 10
radio points available from the very long baseline interferometry
(VLBI) observations at 15 GHz made from 1992 to 2002 using the VLBA
(Very Long Baseline Array). Although the radio sampling was poor we
found a correlation at a confidence level of 90\% between variations
of the optical continuum and radio flux density of the stationary
component (S1) of the jet \citep{arshakian_arshakian08}. The potential source
of optical emission was identified with S1 component - the VLBI core
of the jet located at a distance of $\approx 0.4$ pc from the
central nucleus. Inspired by a relatively high significance of the
correlation (while using only 10 radio points) and the link found
between ejection epochs of jet components and maxima in the optical
light curve, we propose a dense monitoring of six radio galaxies
(including 3C\,390.3) from 2005 using the VLBI and optical
spectroscopy (with $2$\,meter class telescopes in Mexico) to check
previous results.



\section{Structure and kinematics of the VLBI jet}

\begin{figure}[]
\resizebox{\hsize}{!}{\includegraphics[clip=true]{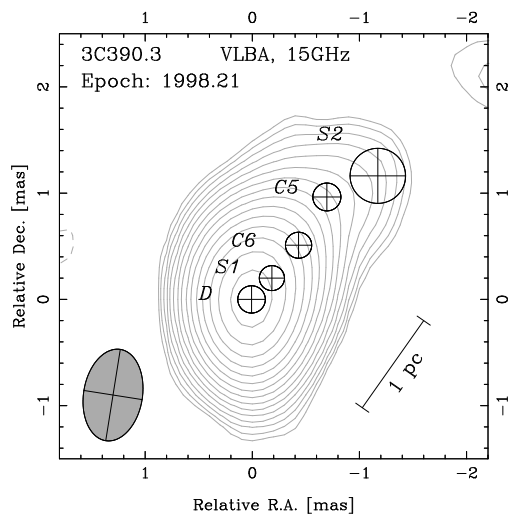}}
\caption{ \footnotesize A single epoch (1998.21) radio structure of
3C\,390.3 observed with VLBA at 15 GHz. }
\label{arshakian_fig:map}
\end{figure}

\begin{figure*}[t!]
\resizebox{\hsize}{!}{\includegraphics[angle=0,clip=true]{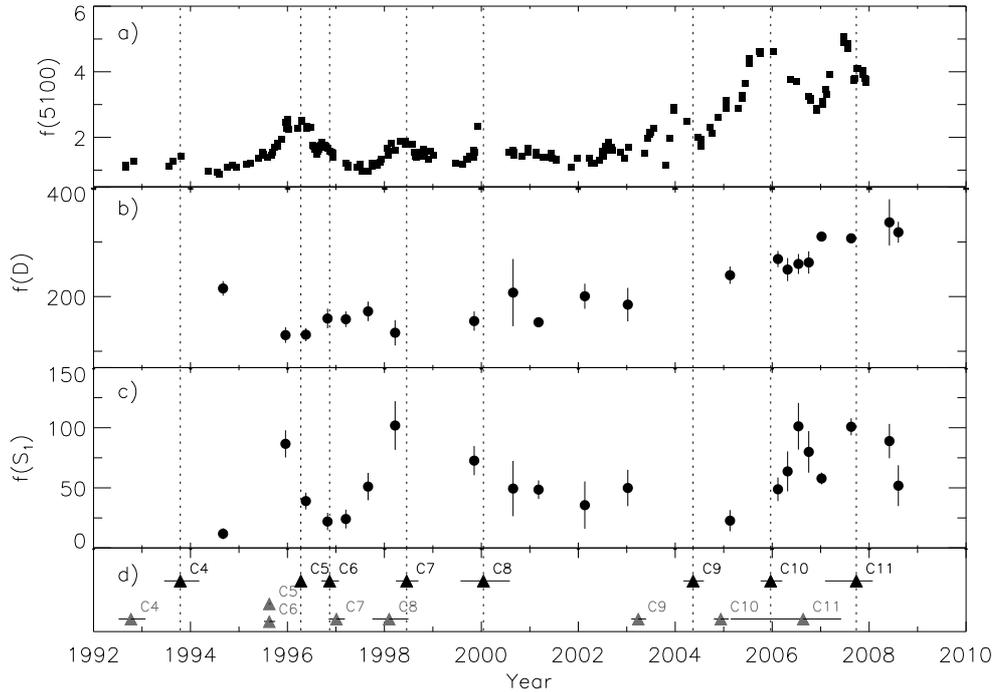}}
\caption{\footnotesize The variations of the optical and radio
fluxes of 3C\,390.3 from 1992 to 2008 \citep[][Shapovalova,
priv. comm.; Tavares et al., in
prep.]{arshakian_shapo01,arshakian_sergeev02}. Optical continuum fluxes at
5100\AA\ (top panel; normalized to $10^{-15}$ erg s$^{-1}$ cm$^{-2}$
\AA$^{-1}$), radio flux densities from D and S1 stationary
components of the jet at 15\,GHz (panels (b) and (c)), and times of
ejection of radio components at D ($t_{\rm D}$; grey triangles in
the panel (d)) and times of passing the moving components through S1
($t_{\rm S1}$; black triangles). }

\label{arshakian_fig:flux-time}
\end{figure*}

Modelfitting of a single epoch of VLBA image (Fig. \ref{arshakian_fig:map})
reveals five radio components on scales of 2 mas (1 mas = 1.09 pc).
For 19 VLBA data sets from 1994 to 2008 we find two stationary
components (S1 and S2) separated from D by 0.28\,mas and 1.5\,mas,
respectively, and eight moving components (C4-C11). Apparent speeds
of moving components lie between 0.8\,c and 1.5\,c. Back
extrapolations of the fits to moving components allows the epochs of
their ejection from D and epochs of passing through the stationary
component S1 to be estimated (see bottom panel in
Fig.~\ref{arshakian_fig:flux-time}). Variations of flux densities of two
innermost stationary components (D and S1) of the jet are presented
in Fig.~\ref{arshakian_fig:flux-time} (panels (b) and (c)).

\section{Couplings between subpc-scale jet and central engine}

\subsection{Jet - optical continuum}
Variations of optical continuum fluxes at 5100\,\AA, H${\rm\beta}$
emission line fluxes, flux densities from D and S1 stationary
components of the jet at 15\,GHz are shown in
Fig.~\ref{arshakian_fig:flux-time}. Both radio (from D and S1) and optical
variations have two independent components: long-term ($\la 2$ yr)
and very long-term ($\ga10$). This is evident also from the
historical light curve of optical continuum in 3C\,390.3 (see, for
example, Shapovalova et al. 2001) which shows a superposition of
short-term flares (few months), long-term (few years) and very long
variations of tens years. Long-term flares from S1 component vary on
timescales of $\la 2$ yr, very similar to optical flares having
comparable timescales (Fig.~\ref{arshakian_fig:flux-time}; panels (a) and
(c)). Moreover, the relative amplitudes of optical and radio flares
are comparable changing the intensity by a factor of 3 for optical
continuum flux and factor of 4 in the case of radio flux from S1.

\begin{figure}[]
\resizebox{\hsize}{!}{\includegraphics[clip=true]{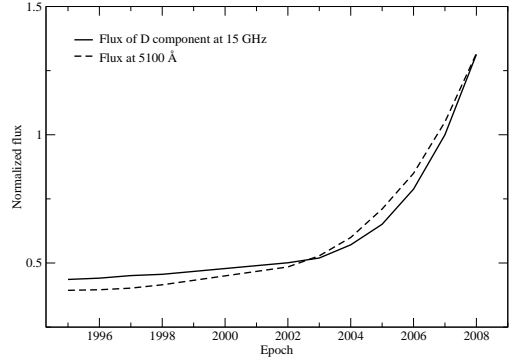}}

\caption{ \footnotesize Fits of normalized optical and radio (from
D) light curves (dashed and solid lines respectively). }

\label{arshakian_fig:nflux_time}
\end{figure}

On the other hand, variations of radio emission from D is more
smooth and follows the changes in the optical continuum on
timescales of $\ga10$ yr (Fig.~\ref{arshakian_fig:flux-time}; panels (a) and
(b)). To demonstrate the similarity of changes in optical and radio
fluxes we fit the radio light curve and selected ranges in the
minima of optical curve by polynomial function of the 4th order and
presented the scaled fits in Fig.~\ref{arshakian_fig:nflux_time}. The optical
and radio (D) fluxes increase by the same factor ($\approx$3 times)
from 1996 to 2008 and have a surprisingly similar behavior. This
suggests that optical variations on timescales of decades are
related to radio variability of the component D, while optical
long-term flares are coupled with radio flares of S1 component of
the jet.

Z-transformed correlation function \citep{arshakian_alexander97} is used to
calculate the correlation between $f$(5100\,\AA) and $f$(S1) as this
method can deal with sparsely sampled data. The distribution of time
lags peaks around $1^{+0.06}_{-0.04}$ year with correlation
coefficient of $0.54^{+0.26}_{-0.22}$. Although the uncertainties of
the correlation coefficient are large (because of small radio
sampling) there is an indication that the correlation could be real.
This is evidenced by a similarity of the relative amplitudes and
timescales of optical and radio (from S1) flares: optical and radio
fluxes rise by a factor of $\approx3$ on timescales of 1.5-2 years.
One could suggest that the radio emission from S1 component leads
the optical continuum emission with time lag of about one year.

We confirm the link between $t_{\rm S1}$ and maxima in the optical
continuum flux \citep[previously reported by][]{arshakian_arshakian08} for two
new components, C9 and C10, ejected around 2004.3 (the C11 component
is not taken into account because of large errors in $t_{\rm S1}$).
All seven ejection events occur within $\sim$0.3 yr after a local
maximum is reached in the intensity of the optical continuum. The
probability that it happens by chance is very small ($<\,$0.0001)
suggesting that the process of passing of radio knots through S1 and
reaching the maxima in the optical intensity are related.


\subsection{Jet - BLR}
\begin{figure}[]
\resizebox{\hsize}{!}{\includegraphics[clip=true]{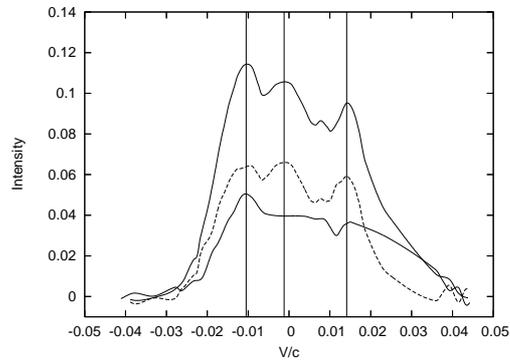}}

\caption{ \footnotesize The averaged H$\beta$ broad-line profiles in
the maximum and minimum states of the optical continuum (top and
bottom solid lines) and the residual profile (dashed line) between
them. }

\label{arshakian_fig:bel}
\end{figure}

The BLR in 3C\,390.3 appears to be complex \citep{arshakian_sergeev02},
probably because of multiple component structure of the emitting
region. To check whether the H$\beta$ broad-line profile changes
during long-term flares of the optical continuum emission, we
averaged the H$\beta$ profiles during the minimum and maximum of the
optical continuum (around 1994.5 and 1995 respectively). The
averaged line profile in the minimum state has two peaks (red and
blue) (Fig.~\ref{arshakian_fig:bel}; bottom solid profile), while in the
maximum state it appears to have an additional central peak
(Fig.~\ref{arshakian_fig:bel}; top solid profile) which is clearly seen also
in the residual profile (dashed profile). The blue and red peaks are
shifted at around $-3000$ km s$^{-1}$ and $+4200$ km s$^{-1}$, and
central peak tends to be blueshifted around $-350$ km s$^{-1}$. This
is a clear evidence that a new broad line region moving towards
observer is excited during the maximum state of the optical flare.


\section{Identification of D and S1 components}
\emph{Identification of the regions D and S1} is important for
understanding the structure of the central engine in 3C\,390.3 and
location of continuum sources producing the variable continuum
radiation \citep{arshakian_arshakian08}. If D would be the base of the
counterjet we should expect that $f$(D)$<f$(S1) because of
relativistic (Doppler) deeming effect. The fact that $f$(D)$>f$(S1)
over 14 years of monitoring period rules out D being the base of the
counterjet. It is likely that D is the base of the jet located near
the BH. This idea is supported by the link between the ejection
epochs of the components C5 and C8 from D and the dip in the X-ray
flux and hardening of the spectrum \citep{arshakian_arshakian06}, similar to
radio galaxies 3C\,120 and 3C\,111 \citep{arshakian_marscher02,arshakian_marscher06}.
This correlation was interpreted as an accretion of the X-ray
emitting gas in the disk into BH and ejection of a fraction of the
infalling matter into the jet \citep{arshakian_marscher02}. The base of the
jet (or D component) is likely to be located near the BH in or above
a hot corona. The jet plasma material is then accelerated and
collimated by helical magnetic fields and produces standing conical
shock \citep{arshakian_gomez95} seen at radio wavelength at a distance of $\la
1$\,pc from the BH. Such a standing feature is seen in VLBI images
of almost all radio-loud AGN and is known as the core of the jet. We
associate the stationary component S1 (at a 0.3\,pc projected
distance from D) with the VLBI core of the jet. \\

\section{Models of the central subpc-scale region}

Here we discuss the results of previous sections and draw possible
emission models of the central engine.

The conventional \emph{source of thermal variable optical emission}
is located in the disk and powered by the accretion of disk material
onto the central BH. Optical radiation ionizes the central BLR
(CBLR) rotating around the BH at a distance $\la 1$ pc. However, in
powerful radio galaxies a \emph{synchrotron variable optical
radiation} can be generated also in the relativistic jet. Depending
on location of the source of optical continuum emission different
regions of broad-line emission can be excited along the jet provided
there is enough gas. According to present models of relativistic
jets \citep[][and references therein]{arshakian_marscher08} the plasma
material accreted onto central nucleus is accelerated and collimated
by helical magnetic fields up to a distance of $\sim 10^4$
Schwarzschild radii. In this region the beamed X-ray, UV, optical
and radio continuum emission can be generated. Due to synchrotron
self absorption in the jet the beamed X-ray emission is radiated
near the central nucleus, while UV and optical emission would appear
with some time delay at larger distances, $\la 1$ pc. The observed
H$\beta$ emission line of the CBLR can be excited by both the beamed
synchrotron optical emission of the jet and thermal optical emission
from the disk. At the end of the acceleration and collimation zone
the jet flow becomes turbulent (because of absence of helical
magnetic fields) and ends in a standing conical shock which is
associated with the VLBI core (S1 component for 3C\,390.3).
Electrons accelerated to high energies in the shock region generate
synchrotron emission from radio to optical bands as evidenced from
multiwaveband polarization variability in the quasar PKS
0420-014 \citep{arshakian_darcangelo07}. The variability of radio and optical
emission from the jet core region can be understood in terms of
disturbances in the jet (moving radio knots) originated near the
central nucleus and spiraling along the jet streamline. Energy
released from interaction of moving radio features with the standing
shock generates the variable non-thermal radio and optical emission,
which can ionize the gas in the rotating subrelativistic outflow
\citep{arshakian_murray97,arshakian_proga00} surrounding the jet and thus ionize the
second outflowing BLR (OBLR) elongated along the jet. In BL Lac type
objects (in which the jet oriented near the line of sight) moving
components also may enhance the radio emission, optical, X-ray and
even Gamma-ray emission when passing near the line of sight of the
observer. In the case of 3C\,390.3 the jet is inclined at $\approx
50^{\circ}$, hence, the Doppler beaming of optical emission from
moving components should be weak but it may ionize the BLR and thus
contribute to the flux of H$\beta$ emission line.

Variations of the H$\beta$ emission line follows the optical
continuum flares with a time lag in the range between $\approx (20$
to $100)$ days \citep{arshakian_dietrich98,arshakian_shapo01,arshakian_sergeev02}. An important
question is in which BLR (CBLR or OBLR) the DP profile is generated.
In 3C\,390.3, the blue and red wings of the H$\beta$ emission line
occur simultaneously \citep{arshakian_shapo01} implying that the DP emission
lines should be generated in the same BLR, either in CBLR or OBLR.
Moreover, the separation between blue and red peaks of the H$\beta$
emission line (Fig.~\ref{arshakian_fig:bel}) anticorrelates with intensity of
optical continuum emission \citep{arshakian_shapo01}.

\subsection{The DP profile generated in the outflowing BLR}
In the model where the non-thermal variable optical continuum
emission of the jet ionizes the DP emission line, the source of
long-term optical continuum emission should be located at a distance
of 0.3\,pc from the radio core S1 (because of time lag of about one
year between radio and optical flares), and at a distance of $\sim
0.6$ pc from the central BH. Such long-term optical flares can be
produced by inverse Compton scattering of radio photons (from S1
region) by relativistic electrons of the jet. Optical flare can
illuminate the outflowing conical BLR giving rise to the DP emission
line profile. Depending on the orientation of the jet the
approaching and rotating outflow should imprint a prominent
signatures on the broad emission lines such as blue-shifted and
single-peaked. The blueshifted central peak of the H$\beta$ profile
(Fig.~\ref{arshakian_fig:bel}) is a signature of the outflowing BLR. Adopting
an inclination angle of $50^{\circ}$ of the jet \citep{arshakian_arshakian08}
we estimate the outflow speed of the BLR to be around $500$ km
s$^{-1}$ while the maximum rotation velocity is $\sim 2000$ km
s$^{-1}$. Very long-term changes of thermal optical continuum
emission are generated in the accretion disk, and correlated with
radio emission of D component located above the disk near the hot
corona.

\subsection{The DP profile generated in the central BLR}
There are two basic ways for attempting to avoid the necessity of
forming at least a fraction of broad-line emission at a large
distance above the accretion disk. In these cases, the broad-line
emission would be produced in a canonical BLR above the accretion
disk, relaxing the need to have a relation between broad-line
emission and the continuum emission from the jet. These two
alternative scenarios are considered below in more detail. The first
alternative is to assume that the long-term and very long-term
variable optical continuum is produced in the vicinity of the
central engine located in D, followed by variations in the radio
regime, with some delay between the two. This scheme requires that
the variable optical continuum is produced near the location of the
component D, and the broad-line emission is generated in a CBLR
above the accretion disk. The radio emission from D should follow
the optical continuum with some delay, which cannot be measured from
the existing data. In the maxima of optical emission the jet
contribution to the continuum becomes significant and manifests
itself by ionizing a conical volume of CBLR and/or OBLR along the
jet direction and producing the central SP component in the H$\beta$
emission line profile (Fig.~\ref{arshakian_fig:bel}). In this scenario, the
correlation observed between the radio and optical continuum fluxes
(Fig.~\ref{arshakian_fig:flux-time}), in which the radio emission leads the
optical continuum emission, has to be ignored. While the relative
sparsity of the radio measurements leaves a theoretical possibility
for such a confusion, the existing data indicate that the
probability of it to happen is less than 5\,\%. Therefore, this
scheme also can hardly be adopted at present.

\subsection{A binary black hole}
The second alternative is to reconsider the binary black hole
scenario proposed for 3C\,390.3 by \cite{arshakian_gaskell96} and discussed by
\cite{arshakian_eracleous97} and \cite{arshakian_shapo01}. In the second scenario, the
double-peaked lines are formed in a binary black hole system in
which each black hole has an independent BLR \citep{arshakian_gaskell96}. The
binary black hole scenario cannot explain the correlated,
simultaneous variability of the red and blue wings of the H$\beta$
line \citep{arshakian_shapo01}. If the two black holes are located in D and
S1, the total mass of the binary must exceed $\sim 5\times 10^9\,
(P_\mathrm{obs}/1000\,\mathrm{yr})\,\mathrm{M}_\odot$. The observed
radial velocity changes indicate a possible periodicity shorter than
100\,yr \citep{arshakian_shapo01}, pushing the total mass of the binary to
$\sim 10^{11}\,\mathrm{M}_\odot$. The trends observed in the radial
velocities cannot be reconciled with the orbital motion in a binary
black hole \citep{arshakian_eracleous97}. Thus a binary system with two black
holes located in D and S1 also has significant difficulties with
explaining the observed properties of 3C\,390.3. The possibility
that the black hole associated with D has a companion with a much
smaller orbital separation cannot be ruled out at present, but it
would face the same difficulties as in the scenario in which the
variable optical continuum is generated in the vicinity of D.\\

\section{Discussion and conclusions}

The scheme proposed in the previous section, where the
long-term variable optical continuum emission is generated in the
jet and DP emission line are produced in the jet-excited BLR,
explains the correlations observed between the radio, optical and
X-ray light curves, and relates them to the evolution of the compact
relativistic jet on scales of $\la 0.5$\,pc. It requires that at
least a fraction of the broad-line emission in 3C\,390.3 is excited
by the non-thermal variable continuum emission from the jet. The
large distance of the jet-excited OBLR from the central engine
challenges the existing models in which the broad-line emission is
localized exclusively around the disk or near the central engine
\citep{arshakian_peterson02}. The existence of the jet-excited OBLR in
3C\,390.3 will question the assumption of virialized motion in the
BLR \citep{arshakian_kaspi00} of all radio-loud AGN, galaxies and quasars,
and, hence, the applicability of the reverberation mapping
\citep{arshakian_peterson02} to estimate the black hole masses of radio-loud
AGN. Time delays and profile widths measured during periods when the
jet emission is dominant may not necessarily reflect the Keplerian
motion in the disk, but rather trace the rotation and outward motion
in an outflow. This can result in large errors in estimates of black
hole masses made from monitoring of the broad-line emission.  In the
case of 3C\,390.3, the black hole mass ($2.1\times10^9$ solar
masses, $M_{\odot}$) estimated effectively from the measurements
near the maximum in the continuum light curve \citep{arshakian_shapo01} is
significantly larger than the values ($3.5$--$4 \times
10^8\,M_{\odot}$) reported in other works \citep{arshakian_wandel99,arshakian_kaspi00}.
This difference is reconciled by considering the line width and the
time delay between the optical continuum and line fluxes near the
minimum of the continuum light curve, which yields $M_\mathrm{bh} =
3.8\times 10^8\,M_{\odot}$. The possible existence of an
outflow-like region in a number of radio-loud AGN should be taken
into account when estimates of the nuclear mass are made from the
variability of broad emission lines.

We analysed the combined radio VLBI (15 GHz) and optical data of
double-peaked radio-loud galaxy over the period from 1994 to 2008.
\begin{enumerate}

\item The structure of the parsec-scale jet is analysed using 19 epochs of
radio VLBI data. Two stationary components (D and S1) and eight
moving radio features were identified on scales of few parsecs.
Apparent speeds of moving components are estimated to be in the
range from 0.8\,$c$ to 1.5\,$c$. Radio flares of S1 component
happens on timescales of 1-2 years (long-term), while a radio flux
density of D component increases gradually during the time period of
16 years (very long-term).

\item Variations of optical continuum has two components, one changing
on timescales of 1-2 years and one on time-scales of few decades,
similar to radio flares of S1 and D stationary components of the
jet. We found a striking similarity in the \emph{very long-term}
behavior of optical and radio (D) emission indicating that they are
physically linked. Identification of D component with the base of
the jet suggests that both very long-term optical and radio emission
have a thermal origin and are generated in the disk and hot corona
above the disk, respectively.

Cross-correlation analysis showed that \emph{long-term} optical
variations follow a radio flares from S1 component with time lag of
$\approx 1$ year. The correlation coefficient is 0.56 with large
uncertainties resulting from the sparsely sampled radio data.
Identification of S1 component as the radio core of the jet
suggested that the source of long-term optical emission is located
at a distance of 0.6 pc from the central BH, and at 0.3 pc from the
VLBI core of the jet (S1 component).

 The link between component S1 and optical continuum is also
supported by the correlation between the local maxima in the optical
continuum light curve and the epochs at which the moving components
of the jet pass the stationary radio feature S1.

\item A model proposed for the central sub-parsec scale region in 3C\,390.3
may account for the observed radio-optical correlations. According
to this model a slowly changing thermal optical continuum emission
is generated in the disk while the relativistic jet flow generates
flares of optical synchrotron radiation on time scales of 1-2 years
and at a distance of $\sim 0.6$ pc from the central BH. The optical
emission of the jet ionizes the rotating and outflowing BLR which
produces the double-peaked profile of the H$\beta$ emission line.

\end{enumerate}

A denser VLBI radio sampling covering a time period of several years
is needed to confirm the correlation between the sub-parsec-scale
jet and variable optical continuum and further to test the model
suggested for the nuclear region of radio-loud galaxy 3C\,390.3. To
understand whether this correlation is common for other galaxies we
started the coordinated long-term radio-optical observations of
nearby radio-loud galaxies.

\begin{acknowledgements}
VHC is supported by CONACYT research grant 54480 (M\'exico).
LCP is supported by Ministry of Science of Serbia (project 146002)
and the Alexander von Humboldt foundation. JT acknowledges support
from the CONACyT program for PhD studies, the International Max
Planck Research School (IMPRS) for Radio and Infrared Astronomy at
the Universities of Bonn and Cologne and the DAAD for a short-term
scholarship in Germany. AIS acknowledges support from RFBR (grant
06-02-16843). The National Radio Astronomy Observatory is a facility
of the National Science Foundation operated under cooperative
agreement by Associated Universities, Inc.

\end{acknowledgements}

\bibliographystyle{aa}

\begin{thebibliography}{}

\bibitem[Alexander(1997)]{arshakian_alexander97} Alexander, T.\ 1997,
Astronomical Time Series, 218, 163

\bibitem[Arshakian \& Belloni(2006)]{arshakian_arshakian06} Arshakian T.G.,
 Belloni T., 2006, VI Microquasar Workshop: Microquasars and Beyond,
  ed Belloni, T., PoS (MQW6), 29

\bibitem[Arshakian et al.(2008)]{arshakian_arshakian08} Arshakian, T.~G.,
Lobanov, A.~P., Chavushyan, V.~H., Shapovalova, A.~I., \& Zensus,
J.~A.\ 2008, Relativistic Astrophysics Legacy and Cosmology -
Einstein's, 189 (arXiv:astro-ph/0602016)

\bibitem[D'Arcangelo et al.(2007)]{arshakian_darcangelo07} D'Arcangelo, F.~D.,
et al.\ 2007, \apjl, 659, L107

\bibitem[Dietrich et al.(1998)]{arshakian_dietrich98} Dietrich, M., et al.\
1998, \apjs, 115, 185

\bibitem[Eracleous et al.(1997)]{arshakian_eracleous97} Eracleous M., Halpern J.P.,
Gilbert A.M., Newman J.A., Filippenko A.V.,\ 1997, ApJ, 490, 216

\bibitem[Gaskell(1996)]{arshakian_gaskell96} Gaskell C.M.,\ 1996, ApJ, 464, L107

\bibitem[G\'omez et al.(1995)]{arshakian_gomez95} G\'omez J.L., Marti,
J.M.A., Marscher A.P., Ibanez J.M.A., Marcaide J.M.,\ 1995, ApJL,
449, L19

\bibitem[Kaspi et al.(2000)]{arshakian_kaspi00} Kaspi S. et al.,\ 2000,
  ApJ, 533, 631

\bibitem[Marscher et al.(2002)]{arshakian_marscher02} Marscher A.P. et
  al., \ 2002, Nature, 417, 625

\bibitem[Marscher(2006)]{arshakian_marscher06} Marscher A.P., 2006, VI Microquasar
  Workshop: Microquasars and Beyond, ed Belloni, T., PoS (MQW6), 25

\bibitem[Marscher et al.(2008)]{arshakian_marscher08} Marscher, A.~P., et
al.\ 2008, \nat, 452, 966

\bibitem[Murray et al.(1997)]{arshakian_murray97} Murray N., Chiang G.,\ 1997,
  ApJ, 474, 91

\bibitem[Peterson et al.(2002)]{arshakian_peterson02} Peterson B.M., 2002,
  Advanced Lectures on The Starburst-AGN Connection, eds Aretxaga I.,
  Kunth D. \& M\'ujica R., Singapore World Scientific, 3

\bibitem[Proga et al.(2000)]{arshakian_proga00} Proga D., Stone J.M., Kallman,
  T.R.,\ 2000, ApJ, 543, 686

\bibitem[Sergeev et al.(2002)]{arshakian_sergeev02} Sergeev S.G., Pronik V.I.,
  Peterson B.M., Sergeeva E.A., Zheng W., \ 2002, ApJ, 576, 660

\bibitem[Shapovalova et al.(2001)]{arshakian_shapo01} Shapovalova I.A. et
  al., \ 2001, A\&A, 376, 775

\bibitem[Ulrich et al.(1997)]{arshakian_ulrich97} Ulrich M., Maraschi L., Urry
  C.M. \ 1997, ARAA, 35, 445

\bibitem[Wandel et al.(1999)]{arshakian_wandel99} Wandel A., Peterson B.M.,
  Malkan M.A.,\ 1999, ApJ, 526, 579

\bibitem[Worrall(2005)]{arshakian_worrall05} Worrall D.M., \ 2005, Multiband
  Approach to AGN, eds Lobanov, A.P. \& Venturi, T. Memorie della
  Societa Astronomica Italiana, 76, 28


%
%
%
%
%
%
%
%
%
%
%
%
%
%
%
%
%
%
%
%
%
%
%
%
%
%
%
%

\end{thebibliography}

\end{document}